\documentclass{ieeeaccess}
\usepackage{cite}
\usepackage{amsmath,amssymb,amsfonts}
\usepackage{algorithmic}
\usepackage{graphicx}
\usepackage{textcomp}
\usepackage{comment}
\usepackage{hyperref}
\usepackage{caption}

\def\BibTeX{{\rm B\kern-.05em{\sc i\kern-.025em b}\kern-.08em
    T\kern-.1667em\lower.7ex\hbox{E}\kern-.125emX}}
\begin{document}
%\history{Date of publication xxxx 00, 0000, date of current version xxxx 00, 0000.}
\doi{}

%\title{Preparation of Papers for IEEE ACCESS}
\title{Enhanced Semantic Graph Based Approach With Sentiment Analysis For
User Interest Retrieval From Social Sites}
\author{\uppercase{USAMA AHMED JAMAL}\authorrefmark{1,2}
}

\address[1]{Glamoureaxdesigns Westheimer Rd, ste M Houston Texas 77082 United States (usama@glamoureaxdrapes.com)}
%\address[1]{Upgrade LZ LLC, Digital Growth Agency, Dubai 25315 United Arab Emirates (e-mail: usama@upgrade-now.com)}
\address[2]{Department of Computer Science and Engineering, HITEC University, Taxila 47080 Pakistan}

\tfootnote{This research was conducted as part of Master Thesis in Computer Science by the first author at HITEC University Taxila.}

\markboth
{Usama Ahmed Jamal }
{Usama Ahmed Jamal }

%\corresp{Corresponding author: First A. Author (e-mail: author@ boulder.nist.gov).}
\corresp{Corresponding author: Usama Ahmed Jamal (e-mail: samusamaa@gmail.com)}

\begin{abstract}

Blogs and social networking sites serve as a platform to the users for expressing their interests, ideas and thoughts. Targeted marketing uses the recommendation systems for suggesting their services and products to the users or clients. So the method used by target marketing is extraction of keywords and main topics from the user generated texts. Most of conventional methods involve identifying the personal interests just on the basis of surveys and rating systems. But the proposed research differs in manner that it aim at using the user generated text as a source medium for identifying and analyzing the personal interest as a knowledge base area of users. Semantic graph based approach is proposed research work that identifies the references of clients and users by analyzing their own texts such as tweets. The keywords need to be extracted from the text generated by the user on the social networking sites. This can be made possible by using several algorithms that extracts the keywords automatically from the
available content provided by the user. Based on frequency and degree it ranks the extracted keywords. Furthermore, semantic graph based model assists in providing useful suggestions just by extracting the interests of users by analyzing their contents from social media. In this approach graph comprises of nodes and edges where nodes represents the keywords extracted by
the algorithm and edges shows the semantic connection between the nodes. The method does not require internet related user activities like surveys or ratings to gather user interest related information.

\end{abstract}


\titlepgskip=-15pt

\maketitle

\section{Introduction}
\label{sec:introduction}

In the current era, the use of internet is increasing more rapidly as the amount and complexity of information and web documents processed on internet are growing day by day. So, it becomes difficult for user to extract the useful information from huge source of web content. Filtering is required for efficient information retrieval. For searching the documents from web search engines or extracting any useful information from documents, or social sites keywords are preferred as they assist in summarizing the concept of document. But in the current era, the semantics technology is considered to be the advanced version that gives relevant results according to the queries and preferences of the users. The intention of the semantic based search is to prioritize the search results based on the semantics of the query and the content. They retrieve efficient results and save the user time and improve the efficiency of system. The use of social networking sites is rapidly increasing and the customers and companies can used it as a platform for interaction. Users may share their common interests using social sites. The significant results according to the preferences of users from social platforms can be achieved and gathered by analyzing the content of ideas expressed by the user on their social sites. 

The problem area is that the system must provide the information regarding the personal interest of users. Currently, researchers are concentrating much more on the recommending and rating systems as well as on personal service systems. There are wide ranges of mini-blogs, shopping, music, movies, and news related sites that recommend their products and updates to the users. For business purposes the use of social networking sites are tremendously evolving. All of the restaurants, hotels gain the popularity through recommendation and personalized systems. Most of the systems recommend their products or services based on their previous search history and purchased items history. But the problem or question arise is that either the system is providing the knowledge related to the users interest or according to their own considerations.

Most of the existing recommending systems suggest useless services to the users just because of their poor knowledge area. Self-interest of end-user is essential requirement for any good recommending system and for that purpose the knowledge about users interests is required. Traditional systems emphasize on gathering the information by surveys, ratings, and tracking users behavior and they are not concerned with the text or content the user shares or updates on their social sites or blogs. The proposed research is concerned with identifying the areas that are of users interest. For extracting the topic related to users interest multi-purpose automatic topic
indexing algorithm is suitable. A social connection serves as a key concept that assists in extracting user interest from social networks. Graphs are used in which users are shown with nodes and edges represent the connection between the users. But considering the semantic similarity of edges is mandatory and useful concept for semantic based search. To generate user interest graph the proposed technique here is semantic graph based approach. The main contributions are as follow:

{\small
\begin{enumerate}

\item  To extract the user interest from the content or social sites according to their preferences, user interest models are generated. This assists in improving the accuracy in a manner that user will acquire the accurate search results, products or services.
\item For keyword extraction using Multi-purpose automatic topic indexing (Muai) algorithm would be beneficial in a manner that it provides the features of increase phrase length, node degree, and semantic relatedness as compared to RAKE algorithm used in baseline approach which has certain limitations of accuracy.
\item Zemanta used serve as web service that also links to DBpedia. It provides more accuracy and used for data (extracted keyword tweets) analysis purpose and serves as text analysis tool too. This would be helpful in extracting the information related to users interest. Sentiment analysis introduce in proposed research depicts the feature in which opinion score used will determine the positive score or negative score of the product according to the users interest.

\end{enumerate}

}

\section{Literature Review}
{
L.M. Jose et al. \cite{1_7684118} proposed a technique based on
semantic graph generation on the basis of users interest.
By analyzing the content of users tweet the preferences
are gathered. Product based graph is also generated and
by matching the user interests graph and product graph the
outcomes are predicted according to the users preferences.
The algorithm used for keyword extraction in this approach
is RAKE and the DBpedia is used for Wikipedia links. W. Yuezhong, et al. \cite{2_7757080} presented the technique for obtaining
high quality data according to users preferences from huge
source of information. In this approach user interest model
is combined with ontology semantic tree and vector space
model. For updating the users interest forgetting function
is used. The architecture of user interest model based on
VSM and ontology comprises of three layers: input layer,
processing layer and output layer. The input layer is primarily
concerned with data acquisition; processing layer is concerned with preprocessing and data processing mechanisms
whereas the output layer represents the user interest model.
A. Akay et al. \cite{3_7464265} presented weighted network model and
applied the technique on health social network site.

In order to describe the users interaction and extract the
knowledge content from posts the approach is used and
comprises of use activity, network clustering and module
analysis. R.Nithish et al. \cite{4_6921974} proposed an approach for mining
the opinions or reviews given by the users about the products
feature. For rating the features of product sentiment analysis
is performed and the domain ontology is used to accumulate
the opinions embedded with product feature description.
Basically the model has three main phases. In first step
domain ontology is created and then the relevant reviews
are extracted from twitter. Finally sentiment analysis needs
to be performed. The technique
for extracting the keywords by using the graph based model
comprised of nodes and edges was also studied. Nodes contain the documents
and edges represent the relationship that exists between the
words of documents. S. Adalbjornsson et al. \cite{6_7760618} presented
extension of Gaussian emission probabilistic latent semantic
analysis model which shows improved accuracy than the
prior latent semantic model by providing the recommendations according to the preferences of users.

Furthermore, two types of recommendation systems are discussed content based and collaborative filtering. In content
based the preferences of users analyzed from the meta-data of
items on the basis of items or products they had purchased or
liked in the past. Collaborative filtering predicts the services
based on analyzing the users interaction with the product
items by observing the consumption behavior of users as well
as on neighbors opinion. V. Jawa et al. \cite{7_7576524} presented a model
in which sentiment analysis are conjugated with interest
graphs that assists the users in suggesting their preferences
ranging from to whom they may follow on social networking
site to the decisions taking regarding what products to be buy
online.

K. Selami et al. \cite{8_sellami2015socialnetworksemanticsocial} presented technique for semantic social
recommendation system. In this approach social aspects and
semantic aspects are merged by using the social network
analysis measures and the semantic similarity measures. E.
Aumayr et al. \cite{9_Aumayr_Hayes_2021} investigated and analyzed the correlation among the topic and user behaviors. The latent topic
encompasses sociability, accessibility, controversy and is
performed on several online communities. M. Saraswat et
al. \cite{10_7724944} uses Explicit Semantic Analysis (ESA), Pointwise
Mutual Information (PMI) and Explicit Semantic Analysis
(ESA) as approaches for topic matching matrix to figure
out semantic relatedness for cross domain recommendations.
Based on the reviews and comments of users recommendations and decisions are suggested. The proposed approach
shows better precision for cross domain recommendation as
it involves semantic relatedness then the cosine similarity
measure which only incorporate topic matching. C.H. Tsai
et al. \cite{11_7424125} proposed architecture comprised of three modules
for effective social personal preference information analysis.

Personal Preference Recommendation Interface module
collects preference data, Data Processing Platform Module
performs data collection and Preference Analysis Arithmetic
Module analyzes preferences of data. The approach carries
out the recommendations on the basis of analyzing the users
behavior and preferences on social sites. M.A. Zingla et
al. \cite{12_ZINGLA2015498} presented approaches for tweet contextualization.
The contextualization is concerned with extracting contents
from tweets. Authors used both the statistical and semantic
approaches. Statistical approach uses association mining
rules whereas semantics approach uses knowledge source
Wikipedia to enhance quality of context by obtaining relevant
context. M. S. Bhuvan et al. \cite{13_7148366} presented a semantic
sentiment analysis model which is scalable and flexible.

Many of the Natural Language Processing (NLP) techniques and machine learning techniques are used for extracting statistical and linguistic patterns from texts. These technologies have found their way into medical and health sciences as it is being used for detecting types of depression based on data from social site Twitter \cite{nusrat2024multiclassdepressiondetection}. At this stage such models serve as primilinary diagnosis and should be verified by an domain expert. Machine Learning applications ranges from homogeneous pattern delineation in atmospheric data \cite{Nusrat_2020_app10196878} to segmentation  and regression of remote sensing data \cite{jamal2023data}. Therefore, on the other hand social media provides a vast domain to be explored though it requires further efficiency for approaches being used because of large amount of data.

The model
depends more on grammar is the limitation of that approach
and can be improved by improving the quality of grammars.
Just by defining the semantics related to any specific domain
the model is said to be flexible. Sentiment polarity and
sentiment score for each review are defined by semantic
patterns and thus the model supports the scalability feature.
F. Zarrinkalam et al. \cite{14_7396850} research work is concerned with
extracting active topics from the users tweet within certain
time period and then rate of users inclination towards the
topic are determined.

Identifying the topics based on their semantics from
the tweet and hence determining the user interests is the
problem areas resolved in the approach \cite{nusrat2023emojipredictiontweetsusing}. Mangal et al.
\cite{15_10.1007/978-3-319-42092-9_2} proposed an approach that exploits the combination of
sentiment analysis and classification of tweets. The approach
assists in extracting the interest related topic areas of the user.
Tweets are collected and sentiment analysis is performed
on the tweets to identify the preferences of user. Then the
classification of tweets under certain labels is performed.
Finally the interest of user is attained according to the
users preferences. Sarivastava et al. \cite{16_6822247} presented Two Parse algorithm named Weighted k-Nearest Neighbor Classifier used as product review analysis system. The system
predicts the reviews of the product in effective manner that
either product is good or bad. In addition, it also satisfies
the customer by providing the percentage of positivity that
the product provides to the customer. The system saves the
customers time because in this manner there is no need to
read each of the review manually.
}

\section{Methodology}

Multi-purpose automatic topic indexing algorithm used
for extracting keywords and ranking them based on their
frequency and degree. Further the core phase is user interest graph generator based on semantics and is comprised
of nodes and edges where nodes represents the extracted
keywords and edges shows the semantic connection between
the nodes.

\subsection{Overview of the domain}

\begin{figure}
    \centering
    \includegraphics[width=0.8\linewidth]{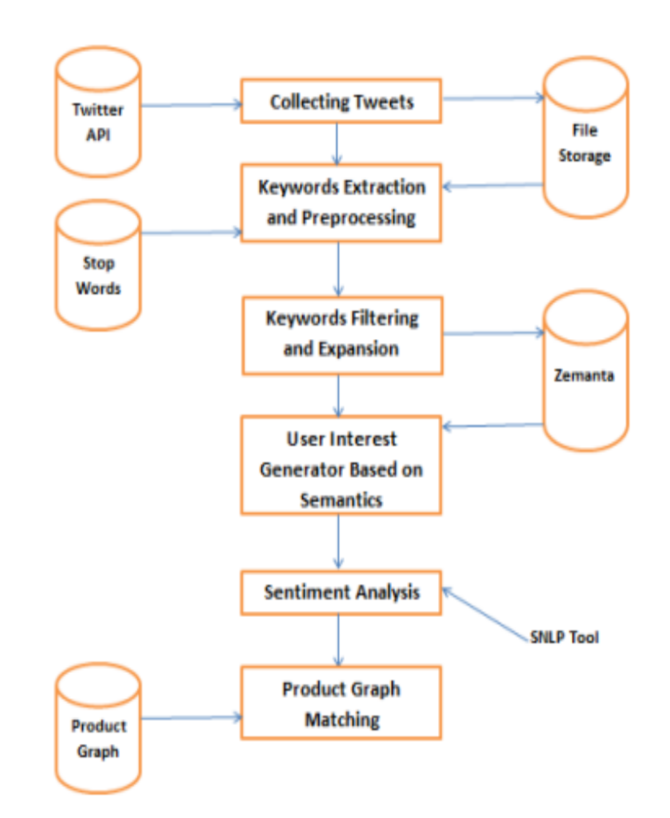}
    \caption{Block Diagram of Methodology}
    \label{figBlock}
\end{figure}

Semantic search and technologies is an advanced version
of existing search mechanism that is primarily concerned
with providing well structured and efficient search results
to users according to their preferences. Web data mining is
actually the domain area of semantic search for extracting the
knowledge of user interest from social media or any other
web document. The data mining concepts are considered for
information extraction and retrieval techniques. The proposed
research work aims at using novel semantic graph based approach. For extracting the interests of users from social sites
Multi-purpose automatic topic indexing (Maui) algorithm is
considered. Figure \ref{figBlock} shows a block diagram for the adopted methodology.

%\begin{figure}
%    \centering
%    \includegraphics{fig1.png}
    %\caption{Figure 1}
    %\label{fig:enter-label}
%\end{figure}

\subsection{Explanation}

In the first phase data collection task is performed in which
Twitter API is used that is concerned with collecting the
tweets of users which they have updated on their accounts.
The tweets in the form of text are stored in files. The stored
tweets act as an input for the next phase. The second phase is
keyword extractions module and is an essential task because
keywords expresses the main topics of the text. During that
phase list of all the keywords are extracted from the users
tweets. These keywords are identified for using in upcoming
phases. The tweets stored in files and the stop words acts
as an input modules for this phase. The stop words include
all the insignificant words that must be filtered out from the
text because they provide lot of unnecessary information.

Keyword extraction involves candidate selection in which
identifying the words or phrases that can be considered
as keywords is the initial activity. Then properties of that
keyword need to be calculated in terms of figuring out
the frequency that how many times the selected candidate
keyword occur, its position in the text and the phrase length.
The keywords are then scored across the tweet. Multipurpose automatic topic indexing (Maui) algorithm is used
for keyword extraction as it provides the features of increase
phrase length, node degree, and semantic relatedness.

Maui serves as pre-processing phase for extraction of keywords from the texts such as tweets. Pre-processing involves
replacing the name of users with real names according to the
screen name as in this way certain ids are allotted to the user.
Zemanta API is used and it includes the links to DBpedia and
provides more accuracy so used in the proposed approach.
Zemanta is concerned with identifying entity from the
users tweet. Inside it is DBpedia that extracts the wikipeida
information. The next phase is keywords filtration, all of the
extracted keywords are checked in Zemanta API and DBpedia that if there exists certain entry related to that keyword or
not. If not then the keywords are removed because they will
not be useful for semantic graph generation. Certain short
descriptions related to list of keywords are added by checking
their corresponding entry in DBpedia inside Zemanta API.
The next phase is user interest graph generator based on
the semantics. The extracted keywords are used as nodes
of graph and the connection between nodes is represented
as edges. The edge is drawn between semantically similar
nodes.

The sentiment analysis of the generated users interest
is then performed by using the SNLP tool. In this phase
opinion score is determined based on three measures positive,
negative and objective. Positive score shows the positive
behavior of word, negative score shows the negative and
objective depicts less subjective behavior of word. Product
graphs are generated from the category of available products
and then product graph matching is performed with the
user interest graph based on their semantics and finally the
outcome shows that the product graphs that relate more to
the users interest are more strongly semantically connected
to the users graph. Connection strength is determined by the
weights that the node possesses. The weights are assigned
to each of the node according to the number of connections
it has with the adjacent node.

\bibliography{bibliography}
%\bibliographystyle{apacite}
%\bibliographystyle{IEEEbib}
%\bibliography{ref}
%\bibliography{bibtex/bib/IEEEabrv.bib,bibtex/bib/IEEEexample.bib}{}
\bibliographystyle{IEEEtran}
\vspace{-15pt}
\EOD
\end{document}